\documentstyle[sprocl,graphicx,array]{article}

\newcommand{\remark}[1]{}

\def\al{\alpha}
\def\be{\beta}
\def\ga{\gamma}
\def\de{\delta}
\def\ep{\epsilon}

\def\ka{\kappa}
\def\la{\lambda}

\def\rh{\rho}

\def\si{\sigma}

\def\ps{\psi}

\def\Ga{\Gamma}

\def\fr#1#2{{{#1} \over {#2}}}
\def\frac#1#2{\textstyle{{{#1} \over {#2}}}}
\def\half{{\textstyle{1\over 2}}}
\def\pt#1{\phantom{#1}}
\def\prt{\partial}

\def\lsim{\mathrel{\rlap{\lower4pt\hbox{\hskip1pt$\sim$}}
    \raise1pt\hbox{$<$}}}
\def\gsim{\mathrel{\rlap{\lower4pt\hbox{\hskip1pt$\sim$}}
    \raise1pt\hbox{$>$}}}

\newcommand{\bequ}{\begin{equation}}
\newcommand{\eequ}{\end{equation}}
\newcommand{\beq}{\begin{eqnarray}}
\newcommand{\eeq}{\end{eqnarray}}
\newcommand{\bea}{\begin{eqnarray}}
\newcommand{\eea}{\end{eqnarray}}
\newcommand{\rf}[1]{(\ref{#1})}

\def\kf{{(k_{F})}}
\def\kkaf{{$(k_{AF})_\mu$}}
\def\kkf{{$(k_{F})_{\ka\la\mu\nu}$}}

\def\a{$a_\mu$}
\def\b{$b_\mu$}
\def\c{$c_{\mu\nu}$}
\def\d{$d_{\mu\nu}$}
\def\e{$e_\mu$}
\def\f{$f_\mu$}
\def\g{$g_{\la\mu\nu}$}
\def\H{$H_{\mu\nu}$}

\newcommand{\eqref}[1]{(\ref{#1})}

\newcommand{\incps}[5]{\includegraphics[#2,#3][#4,#5]{#1}}

\newcommand{\incpicwh}[3]{\includegraphics[width=#2,height=#3]{#1}}

\newcommand{\pad}{\hspace{-0.6cm}}
\newcommand{\fgl}[1]{\hspace{0.65cm}#1\hspace{-0.65cm}}
\newcommand{\qedpic}[1]{\incpicwh{#1}{2.1cm}{2.1cm}}
\newcommand{\negpad}{\!\!\!\!\!\!\!\!}

\begin{document}

\title{LORENTZ VIOLATION AT ONE LOOP}

\author{AUSTIN G.M.\ PICKERING}
\address{Physics Department, Indiana University,
          Bloomington, IN 47405, U.S.A.}
\maketitle

\abstracts{
The proof of one-loop renormalizability of the general Lorentz- and
CPT-violating extension of quantum electrodynamics is described.
Application of the renormalization-group method is discussed and implications for theory and experiment are considered.
}

\section{Introduction}

Invariance of the 
standard model under both Lorentz and CPT transformations is confirmed to high
precision by experiment.\cite{cpt98} However, it is still possible that these
symmetries may be violated in nature by interactions which are too small to be
observable by current experiments.
A framework exists for the description of such effects in terms of a small
perturbation to the standard model at low energies called the standard-model
extension.\cite{ck} This theory, based on a lagrangian which is observer
Lorentz covariant and is constructed from the fields of the standard model,
forms the basis for much of the work described in this volume and its parameters
have been bounded by various
experiments.\cite{cpt98}
It is considered
to be the low energy limit of some more complete theory, valid at the Planck
scale, such as noncommutative field theory\cite{chklo} or string
theory.\cite{kps}

It is of interest to ask how the low energy quantum field theory
relates to the underlying theory at high energies. Studies of
microcausality and stability in the standard model extension\cite{kle} suggest
that nonrenormalizable terms play an
essential role in the theory at energies approaching the Planck scale. The aim
of this talk is to investigate the relationship between the low and
high energy theories from a different perspective. I will describe the results
of applying the renormalization-group method\cite{rg,jc} to the general Lorentz- and
CPT-violating extension of QED. Although some work has been
done to calculate one-loop contributions to this theory, it is
incomplete and so the full one-loop analysis,\cite{klp} including a  proof of
renormalizability at one-loop, a generalization of the Furry theorem\cite{wf} and the
calculation of all one-loop divergences will be described here.

The renormalization-group method is used to study a quantum field theory over
a wide range of energies. In
particular,
the running of the parameters in the theory over the given
energy range must be found by calculating the relevant beta functions, usually
perturbatively. Such is the importance of the beta function
in conventional quantum field theory that it is now known  up to three loops for
a general gauge field theory.\cite{betfun} The
calculation of the eeta functions for the parameters which describe Lorentz- and
CPT-violation is therefore a significant step in
any attempt to understand the standard-model extension and its relation to any
underlying theory. I will describe how the one-loop beta functions
have been used to find the running of the parameters in the theory and the
implications for experiment and theory will be considered.

\section{Framework}
\label{secttheoframe}

The lagrangian $\cal L$ for a fermion field $\ps$ of mass $m$
is \cite{ck}
\bea
{\cal L} &=& \frac{1}{2} i \bar{\ps} \Ga^\mu
\stackrel{\leftrightarrow}{D_\mu} \ps  -  \bar{\ps}
M \ps  - \frac{1}{4} F^{\mu\nu} F_{\mu\nu}
\nonumber \\ & &
- \frac {1}{4}
(k_F)_{\ka\la\mu\nu} F^{\ka\la} F^{\mu\nu} + \frac{1}{2}
(k_{AF})^{\ka} \ep_{\ka\la\mu\nu} A^\la F^{\mu\nu},\label{lagdef}
\eea
where $\Ga^{\nu} = \ga^{\nu} + \Ga^{\nu}_1$
and
$M = m + M_1$,
with
\beq
\Ga^{\nu}_1 &\equiv&
c^{\mu\nu} \ga_\mu + d^{\mu\nu} \ga_5
\ga_\mu + e^\nu + i f^\nu \ga_5 + \half g^{\la\mu\nu}
\si_{\la\mu},
\nonumber \\
M_1 &\equiv& a_\mu \ga^\mu + b_\mu \ga_5 \ga^\mu +
\half H_{\mu\nu} \si^{\mu\nu}.
\label{gamdef}
\eeq

The coefficients \a, \b, \c, \d, \e, \f, \g, \H, \kkaf\ and \kkf\ control
Lorentz violation and are real because $\cal L$ is hermitian.
Of these,
\a, \b, \e, \f, \g\ and \kkaf\ control CPT violation, $M_1$  and \kkaf\
have dimensions of mass,
and
$\Ga_1^\mu$ and \kkf\ are dimensionless.
\c\ and \d\ are traceless,
\H\ and \g\ are antisymmetric
on their first two indices, \kkf\
has the symmetries of the Riemann tensor
and ${(k_F)_{\mu\nu}}^{\mu\nu} = 0$.

The lagrangian in \rf{lagdef} is observer Lorentz invariant by
construction,\cite{ck} but
it is not invariant under particle Lorentz transformations which leave the
coefficients
for Lorentz violation unchanged. Inertial frames exist  where, as a result of 
their observer Lorentz dependence,
these coefficients are very large and a perturbative expansion in them is 
inappropriate. This work is restricted to so-called
concordant frames,\cite{kle} where the parameters that determine the Lorentz
violation are extremely small, such as in the Earth frame. In this context, a
one-loop diagram is therefore defined to contain exactly one closed
loop and to be at most first order in the coefficients for Lorentz
violation.  Note that external
propagators are not used to calculate the effective action, so
it is possible to expand perturbatively in the coefficients for Lorentz 
violation. Problems arise with Hilbert space if one attempts this approximation
with external legs.\cite{ck}

Table 1 lists the C, P, and T transformation properties of the field
operators in \rf{lagdef}, labeled by their associated coefficient. Because QED is
itself invariant under both rotation and CPT transformations, the
above restriction to linear violation of Lorentz symmetry means that, at this
order, the coefficients for Lorentz violation can only receive radiative
corrections from coefficients with exactly the same symmetry properties.

\smallskip

\centerline{\small
Table 1: Discrete-symmetry properties.
}
\begin{tabular}{|c||c|c|c||c|c|c||c|}
\hline
\parbox{3cm}{~~~~~~} & $\rm C$ & $\rm P$ & $\rm T$ &
 $\rm CP$ & $\rm CT$ & $\rm PT$ & $\rm CPT$
\\ \hline   
\parbox{3cm}
{\quad $c_{00}$,$(k_F)_{0j0k}$,
\\ \pt{ccc}$c_{jk}$,$(k_F)_{jklm}$}
  &$+$&$+$&$+$&$+$&$+$&$+$&$+$
\\ \hline
$b_j,g_{j0l},g_{jk0},(k_{AF})_j$
  &$+$&$+$&$-$&$+$&$-$&$-$&$-$
\\ \hline
$b_0,g_{j00},g_{jkl},(k_{AF})_0$
  &$+$&$-$&$+$&$-$&$+$&$-$&$-$
\\ \hline
$c_{0j},c_{j0},(k_F)_{0jkl}$
  &$+$&$-$&$-$&$-$&$-$&$+$&$+$
\\ \hline
$a_0,e_0,f_j$
  &$-$&$+$&$+$&$-$&$-$&$+$&$-$
\\ \hline
$H_{jk},d_{0j},d_{j0}$
  &$-$&$+$&$-$&$-$&$+$&$-$&$+$
\\ \hline
$H_{0j},d_{00},d_{jk}$
  &$-$&$-$&$+$&$+$&$-$&$-$&$+$
\\ \hline
$a_j,e_j,f_0$
  &$-$&$-$&$-$&$+$&$+$&$+$&$-$
\\ \hline
\end{tabular}

\smallskip

For instance, $a_\mu$ has the same symmetries as $e_\mu$ and one might expect 
these parameters to mix. However, in a
mass-independent renormalization scheme, such as dimensional
regularization,\cite{dr1,dr2} divergent radiative corrections are polynomial in
the massive parameters. Thus while $a_\mu$ can receive corrections from $e_\mu$,
it follows that, on dimensional grounds, $e_\mu$ cannot receive corrections from
$a_\mu$. In addition, rotational symmetry prevents $e_0$ and $f_j$ from mixing
in this approximation, despite their matching C,P and T transformation
properties.

\section{Renormalizability at one loop}
\label{renability}

For a quantum field theory to be renormalizable it is required to have a finite
number of divergent, one-particle irreducible Green functions and a sufficient number of
parameters in the theory to absorb these divergences.
Using the following Feynman rules, deduced from the lagrangian
in \rf{lagdef}, it can be shown that there are indeed a
finite number of superficially divergent Green functions contributing to the effective action.

The propagator for a fermion
of momentum $p^\mu$, in
the same direction as the charge arrow shown on the diagram, is
\beq
\raisebox{-0.4cm}{\incps{ferm.eps}{-1.5cm}{-.5cm}{1.5cm}{.5cm}}
 &=& i \fr{(\ga_\mu p^\mu  + m )} {p^2 - m^2 },
\eeq
The $\Ga_1^\mu$ and $M_1$ terms give the following propagator insertions
\beq
\raisebox{-0.4cm}{\incps{fermm.eps}{-1.5cm}{-.5cm}{1.5cm}{.5cm}}
= - i M_1, & &
\raisebox{-0.4cm}{\incps{fermgam.eps}{-1.5cm}{-.5cm}{1.5cm}{.5cm}}
= i \Ga_1^\mu p_\mu.
\eeq
The usual photon propagator, with  gauge fixing parameter, $\al$,
\beq
\mu \negpad \raisebox{-0.4cm}{\incps{phot.eps}{-1.5cm}{-.5cm}{1.5cm}{.5cm}}
\negpad \nu &=& -
\fr{i}{p^2} \Big[ \de_{\mu\nu} - \fr{p_\mu p_\nu}{p^2} (1-\al)
\Big],
\eeq
receives the following additional insertions, due to \kkf\ and \kkaf,
\beq
\mu \negpad
\raisebox{-0.4cm}{\incps{photkf.eps}{-1.4cm}{-.5cm}{1.4cm}{.5cm}}
\negpad \nu =  -
2 i p^\al p^\be {k_F}_{\al\mu\be\nu}, & \,\,\,\, &
\mu \negpad
\raisebox{-0.4cm}{\incps{photkaf.eps}{-1.4cm}{-.5cm}{1.4cm}{.5cm}}
\negpad \nu = 2
k_{AF}^\al \ep_{\al\mu\be\nu} p^\be.
\eeq
In addition to the usual fermion-photon vertex, there is also a vertex
due to $\Ga_1^\mu$. These are given in turn by
\beq
\raisebox{-.5cm}{\incps{vert.eps}{-1.5cm}{-0.1cm}{1.5cm}{1cm}} = -
i q \ga^\mu
&\hspace{1em}\mathrm{and}&
\raisebox{-.5cm}{\incps{vertgam.eps}{-1.5cm}{-0.1cm}{1.5cm}{1cm}}
= -i q \Ga_1^\mu,
\eeq
where $q$ is the fermion charge
and $\mu$ is the space-time index on the photon line.

The superficial degree of divergence $D$
of a general diagram contributing to the effective action  is then
\beq
D &=& 4 - \frac{3}{2} E_\ps - E_A - V_{M_1} - V_{AF},
\label{degdiv}
\eeq
where $E_\ps$ and $E_A$ are the number of external fermion and photon legs
and  $V_{M_1}$ and $V_{AF}$ are the number of $M_1$ and
\kkaf\ insertions, respectively.
The superficially divergent diagrams have $D \ge 0$ and include
the usual divergent diagrams which arise in conventional QED at one loop, (Fig.\ 1).
The remaining one-loop divergent diagrams have the same topologies as Fig.\ 1 but include
exactly one insertion of a Lorentz-violating operator.
For example, Fig.\ 2 shows the set of such diagrams with the same topology as
Fig.\ 1e.
\begin{figure}
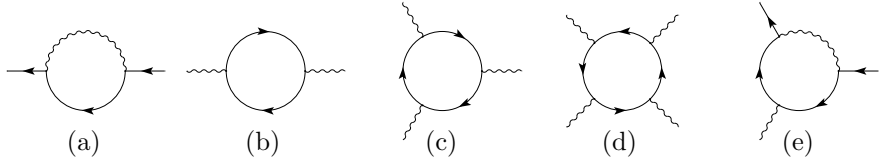

\begin{tabular}{ccccc}
 \fgl{(a)}\pad\qedpic{epqed11.eps}\pad &
 \fgl{(b)}\pad\qedpic{p2qed11.eps}\pad &
 \fgl{(c)}\pad\qedpic{p3qed11.eps}\pad &
 \fgl{(d)}\pad\qedpic{p4qed11.eps}\pad &
 \fgl{(e)}\pad\qedpic{phepqed11.eps}\pad
\end{tabular}
\caption{1PI divergent diagram topologies for QED}
\label{gphs}
\end{figure}

\begin{figure}
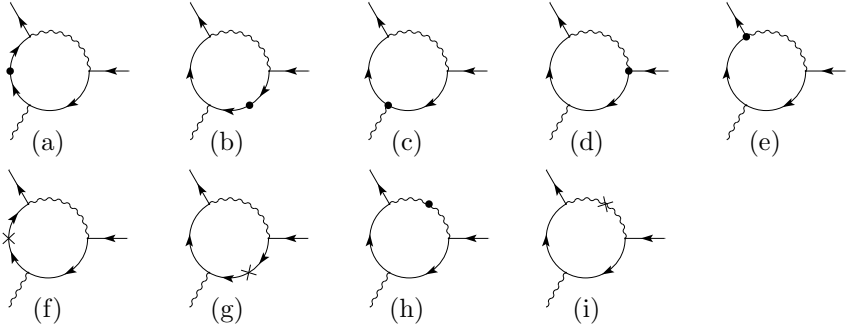

\begin{tabular}{ccccc}
\fgl{(a)}\pad \qedpic{phepgam11.eps}\pad
& \fgl{(b)}\pad \qedpic{phepgam12.eps}\pad
&
 \fgl{(c)}\pad \qedpic{phepgamph11.eps}\pad
& \fgl{(d)}\pad \qedpic{phepgamph17.eps}\pad
&
\fgl{(e)}\pad \qedpic{phepgamph18.eps}\pad
\\ \fgl{(f)}\pad \qedpic{phepmass11.eps}\pad
&
\fgl{(g)}\pad \qedpic{phepmass12.eps}\pad &
 \fgl{(h)}\pad \qedpic{phepkayeff11.eps}\pad &
 \fgl{(i)}\pad \qedpic{phepkaf11.eps}\pad
\end{tabular}
\caption{Divergent fermion-photon diagrams in the QED extension}
\label{cptgphs}
\end{figure}

The divergent integrals are regulated using dimensional regularization in
$4-2\epsilon$ dimensions.\footnote{The analysis has also been performed in the
Pauli-Villars scheme and yields equivalent results.} Working at linear
order in the coefficients for Lorentz violation enables these coefficients to be
taken outside the integral and ensures that all integrands transform in the
conventional manner under particle Lorentz transformations. They can thus be
dealt with by conventional techniques and for similar reasons there are no
problems performing the Wick rotation.

Throughout this work, the naive definition of $\ga_5$ in $d$ dimensions, which
anticommutes with all other $\ga$-matrices, is used. This is  because we are only
interested in the divergent contributions to the effective action and at one loop,
this simplification leads only to irrelevant errors in the finite corrections.

Unlike tree-level calculations, loop calculations involve an integration over an
infinite range of momenta, but because of the difficulties with stability
and microcausality near the Planck scale\cite{kle} the validity of the
Feynman rules at this energy scale is unclear. However, it
is customary to assume that the low energy physics is not sensitive
to the details of the physics at high energy, and therefore it is reasonable to
employ the Feynman rules over the entire range of the integration. Further
justification of this assumption would be of interest.

In an abelian gauge field theory there can be no divergent three- or four-point
photon interactions because they do not correspond to a tree-level interaction.
In conventional QED, absence of the three-point radiative corrections is assured
by the Furry theorem\cite{wf}
 which states that there are in fact two nonzero three-point diagrams with
opposing charge flow that cancel each other precisely. This depends crucially on
the transposition properties of the $\ga$-matrices at the fermion-photon vertex.
The QED extension has more complicated $\ga$-matrix structure and an example of a
situation arising in this more general case is illustrated in Fig.\ 3, where
there is an insertion of a Lorentz violating operator, $\Ga_1^\nu$, at one of the
vertices. The sum of these two diagrams is now proportional to
$( \Ga_1^{\nu} - \tilde\Ga_1^{\nu} )$,
where
\beq
\tilde\Ga^{\nu}_1 &\equiv&
 c^{\mu\nu} \ga_\mu - d^{\mu\nu} \ga_5 \ga_\mu
- e^\nu - i f^\nu \ga_5
+ \half g^{\la\mu\nu} \si_{\la\mu}.
\eeq
and so the $\ga_\mu$ terms cancel as in the usual Furry analysis, as do the
$\si_{\la\mu}$ terms, but the  $I$, $\ga_5$, $\ga_5 \ga^\mu$ do not.
Similar arguments show that the same conclusion is true for $\Ga_1^\mu$
insertions in propagators, but the opposite is true for $M_1$
insertions, where it is the $\ga_\mu$ and $\si_{\la\mu}$ terms which
survive while the others cancel.
The analysis is similar for the four-point vertex, but in this case the sum is
proportional to $(\Ga_1^\mu + \tilde\Ga_1^{\mu})$, which means that
it is now the $\ga_\mu$ and $\si_{\la\mu}$ terms which survive and the $I$,
$\ga_5$, $\ga_5 \ga^\mu$ which cancel. Likewise for $M_1$ insertions it is the
$\ga_\mu$ and $\si_{\la\mu}$ terms which cancel while the others survive. These
arguments are applicable to diagrams with any number of photon legs with linear
insertions of Lorentz-violating operators.

\begin{figure}
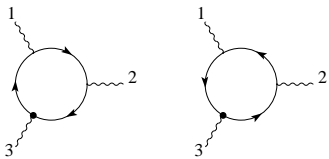

\centering
\begin{tabular}{cc}
  \qedpic{p3gamph11.eps}
&   \qedpic{p3rgamph11.eps}
\end{tabular}
\caption{Two contributions to the cubic photon interaction.}
\label{cptp3gphs}
\end{figure}

It follows immediately that there are no corrections to the three-point photon
vertex proportional to \b, \c, or \g\
and no four-point corrections dependent on \a, \d, \e, \f, or \H.
Other contributions must be explicitly evaluated and there are in fact $n$ of each type for an $n$-point
photon interaction, as illustrated in Fig.\ 4 for a three-point interaction with
$\Ga_1^\mu$ propagator insertion. The sum of all such terms for
a given coefficient is
zero, as required for renormalizability.

\begin{figure}
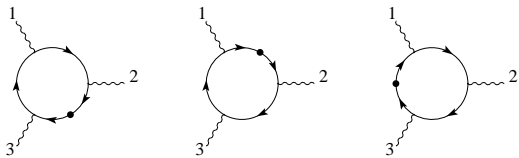

\centering
\begin{tabular}{ccc}
  \qedpic{p3ec1.eps}   &
   \qedpic{p3ec2.eps}  &
   \qedpic{p3ec3.eps}
\end{tabular}
\caption{Permutations for the cubic photon vertex.}
\label{ecgphs}
\end{figure}

The remaining one-loop divergences in the theory\cite{klp} must be removed
by defining renormalization constants in the usual way,
\beq \ps_B &=&
\sqrt{Z_\ps} \ps,\quad A^\mu_B = \sqrt{Z_A} A^\mu,\quad
m_B = Z_m m,
\eeq
as well as renormalizing the coefficients for Lorentz violation such that:
\bea
a_{B\mu} =
(Z_a)_\mu^{\pt{\mu}\al} a_\al,
\quad
b_{B\mu} =
(Z_b)_\mu^{\pt{\mu}\al} b_\al,
\quad
c_{B\mu\nu} =
(Z_c)_{\mu\nu}^{\pt{\mu\nu}\al\be} c_{\al\be} \ldots
\label{bare}
\eea
and so on.
(Bare parameters are denoted with a subscript $B$.)
These factors can be determined\cite{klp} from  the propagator
corrections and the conventional QED fermion-photon vertex alone. They are,
however, sufficient to renormalize all of the divergences which appear at
one-loop order, including those which arise from the extra corrections to the
fermion-photon vertex due to the presence of the Lorentz-violating operators in Eq.\
\rf{lagdef}. This is because the Ward identity of QED, $Z_q\sqrt{Z_A} =1$,  is
preserved, meaning that the Lorentz-violating extension of QED preserves gauge
invariance and is renormalizable at one-loop.

\section{Application of the Renormalization Group}

In a renormalizable quantum field theory one can apply the renormalization-group
technique to study the evolution of the couplings over a wide range of energies.
In the present case we can only be certain that the theory is renormalizable at
one-loop order. In the paper by Kosteleck\'y et al.\cite{klp} there is a 
discussion of the issues concerning the applicability of the usual
renormalization-group techniques to the QED extension of Eq.\ \rf{lagdef}. It is
argued that it is reasonable to assume that one can apply the usual
renormalization-group technique at one-loop in a theory which is
multiplicatively renormalizable to that order. Beyond that order there may be
nonrenormalizable contributions  which would
invalidate the assumption of multiplicative renormalizability used to derive
the renormalization-group equation. The assumption that the
renormalization-group method is applicable at one-loop can be summarized as the
assumption that it is reasonable to ignore higher loop effects in a one-loop
calculation as long as the couplings which describe the physics in this
approximation remain small and any
nonrenormalizable couplings remain negligible. Clarification of the
extent to which this assumption is justified would be of interest.

Bearing in mind these caveats, it is possible to apply the renormalization-group
method to the calculation of the running couplings 
 in a way that is indistinguishable
in practice from the usual approach for an all-orders renormalizable quantum
field theory.
This method relies on a knowledge of the beta functions for the parameters in the theory.
In a theory with couplings $\{x_j\}, j = 1,2,\ldots,N$ the beta function \cite{jc} for $x_j$, is definied as 
\beq
\be_{x_j} &\equiv& \mu \fr{d x_j}{d\mu},
\label{betadef}
\eeq
where $\mu$ is the mass parameter introduced to define the
regularization scheme.

In dimensional regularization, it is usual to renormalize each parameter as follows:
\beq
x_{jB} &=& \mu^{\rh_{x_j} \ep} Z_{x_j} x_j,
\label{baredef}
\eeq
where the introduction of $\mu^{\rh_{x_i}}$
ensures that the bare parameter, $x_{jB}$, and its associated renormalized
parameter $x_j$
have the same dimension.
It is easy to show that, for the QED extension $\rh_q = 1$ and all other
$\rh_{x_i}$ are zero.
The usual analysis\cite{gt} then shows that the beta function is given by
\beq
\be_{x_j} &=&
\lim_{\ep \to 0}\Big[
- \rh_{x_j} a^j_1
+ q \fr{\prt a^j_1 }{\prt q}
\Big],
\label{betacalc}
\eeq
which involves only the simple $\epsilon$-poles, $a_1^j$, in the renormalization
factors $Z_{x_j}$.
Hence the beta functions, and therefore the running of the various parameters in this theory, is determined by the $q$-dependence of the renormalization factors alone.

For example, the renormalization factor for \d\ is\cite{klp}
\beq
(Z_d)_{\mu\nu}^{\pt{\mu\nu}\al\be} d_{\al\be}
= d_{\mu\nu} + \fr{q^2}{12\pi^2 \ep}
( d_{\mu\nu} + d_{\nu\mu} ),
\eeq
and hence, using Eq.\ \rf{betacalc}, the beta function is found to be
\beq
(\be_d)_{\mu\nu} &=& \fr{q^2}{6\pi^2} ( d_{\mu\nu} + d_{\nu\mu} ).
\eeq
From Eq.\ \rf{betadef}, this can be rewritten as
\beq
\fr{d}{d\ln \mu} \Big[ Q^2 (d + d^T) \Big] &=& 0.
\label{deerun}
\eeq
where
\beq
Q(\mu) &\equiv& 1 - \fr{q_0^2}{6\pi^2} \ln \fr{\mu}{\mu_0},
\label{qdef}
\eeq
determines the running of $q$
with the scale $\mu$ in conventional QED. In particular, with the boundary
conditions $x_{j0}  \equiv  x_j(\mu_0)$, the standard result is $q(\mu)^2 =
Q^{-1} q_0^2$ and it follows from Eq.\ \rf{deerun}, that \d\ runs like $Q^{-2}$.

In a similar way the running of the other coefficients for Lorentz violation has
been found.\cite{klp} All of the mixing is consistent with the predictions made
at the end of section \ref{secttheoframe}.
The running is determined entirely by $Q(\mu)$ in Eq.\ \eqref{qdef},
but the powers of $Q(\mu)$ involved range from $-3$ to $9/4$.
For example, \a\ and \c\ run as follows
\beq
a_\mu &=& a_{0\mu} - m_0(1 - Q^{9/4}) e_{0\mu},
\nonumber\\
c_{\mu\nu} &=&
 c_{0\mu\nu}  
- \frac 13 (1-Q^{-3})
( c_{0\mu\nu} + c_{0\nu\mu}
- \kf_{0\mu\al\nu}^{\pt{0\mu\al\nu}\al} ).
\eeq
while \e\ does not run at all in this approximation.
\begin{figure}
\centering
\setlength{\unitlength}{0.0097cm}
\begin{picture}(1055,600)(0,0)
\put(25,592){
{\makebox(0,0){$Q^n$}}
}
\put(88,100){
\makebox(980,500)[r]{\incpicwh{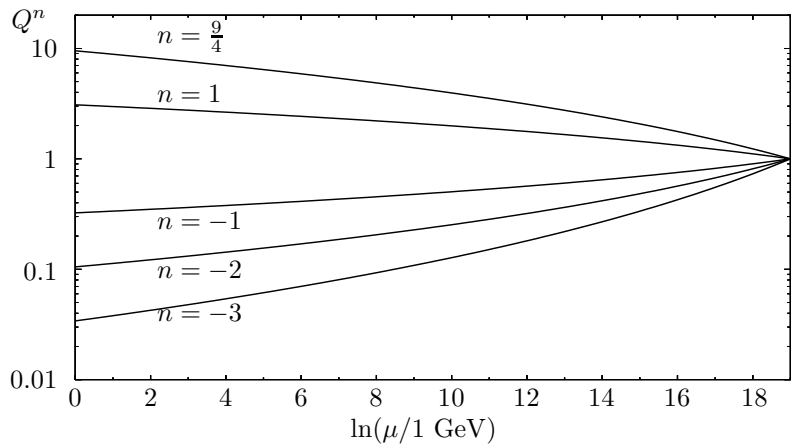}{980\unitlength}{500\unitlength}}
}
\put(75,0){
\makebox(980,60)[c]{$\ln(\mu/\mathrm{1~GeV})$}
}
\put(200,100){
 \begin{picture}(0,0)
  \put(0,90){\makebox(0,0)[l]{$n=-3$}}
  \put(0,150){\makebox(0,0)[l]{$n=-2$}}
  \put(0,217){\makebox(0,0)[l]{$n=-1$}}
  \put(0,392){\makebox(0,0)[l]{$n=1$}}
  \put(0,471){\makebox(0,0)[l]{$n=\frac{9}{4}$}}
 \end{picture}
}
\put(63,100){
 \begin{picture}(0,0)
  \put(0,0){\makebox(0,0)[r]{$0.01$}}
  \put(0,149){\makebox(0,0)[r]{$0.1$}}
  \put(0,302){\makebox(0,0)[r]{$1$}}
  \put(0,454){\makebox(0,0)[r]{$10$}}
 \end{picture}
}
\put(88,75){
 \begin{picture}(0,0)
  \put(0,0){\makebox(0,0){$0$}}
  \put(103,0){\makebox(0,0){$2$}}
  \put(206,0){\makebox(0,0){$4$}}
  \put(309,0){\makebox(0,0){$6$}}
  \put(412,0){\makebox(0,0){$8$}}
  \put(515,0){\makebox(0,0){$10$}}
  \put(618,0){\makebox(0,0){$12$}}
  \put(721,0){\makebox(0,0){$14$}}
  \put(824,0){\makebox(0,0){$16$}}
  \put(930,0){\makebox(0,0){$18$}}
 \end{picture}
}
\end{picture}
\caption{Variation of the function $Q(\mu)^n$.}
\label{runningpic}
\end{figure}

To apply this analysis to a realistic theory one must include the full
standard-model fermion content and possibly allow for the effects of charged
scalars as well as the embedding of the U(1) gauge group in a unification
group. The precise behavior of the function $Q(\mu)$ is determined by the
value of the coefficient of $\ln(\mu/\mu_0)$, which is highly
model-dependent. It is therefore difficult to apply the results of these
calculations directly to the realistic situation, but one might argue\cite{klp}
that a reasonably realistic choice for this coefficient is simply $1$. This particular case
has been plotted in Fig.\ 6 to show the behavior of $Q(\mu)^n$ between the weak
and Planck scales for various values of $n$ which appear in the running
equations for the coefficients for Lorentz violation. The running is
far too small to account for the extreme
suppression of these coefficients if they are assumed to be ${\cal O}(1)$ at the
Planck scale. One might expect nonrenormalizable
terms, known to be essential to maintain stability and causality near the Planck
scale,\cite{kle} to  increase the rate of running of the parameters because
they introduce negative mass dimension couplings which would be expected to run
faster than those considered here.

Figure 6 shows that
the variation of $Q(\mu)$ with the scale $\mu$
is large enough to produce an appreciable spread at low energies in the values
of the various coefficients describing Lorentz violation.
In addition, in the full standard-model extension it would be necessary to
include the effects of the factors $Q_2(\mu)$ and $Q_3(\mu)$
arising from running of the SU(2) and SU(3) gauge couplings,
respectively.
Therefore, because of the potential effects of renormalization-group running
on the relative sizes of the parameters describing Lorentz violation, this work
illustrates the importance of placing experimental bounds  on  coefficients from
all sectors of the standard-model extension and not simply assuming that they
are of the same order of magnitude at the low energy scale.

\section*{Acknowledgments}
I would like to thank Alan Kosteleck\'{y} and Chuck Lane for their
collaboration and John Gracey and Tim Jones for helpful discussions. This work
was supported in part by the United States Department of Energy
under grant number DE-FG02-91ER40661.

\end{document}